\documentclass[a4paper,12pt]{article}
\usepackage{amsmath,amsfonts,amssymb,cite}

\begin{document}

\begin{center}
{\LARGE\bf Lower and upper bounds on the mass of light
quark-antiquark scalar resonance in the SVZ sum rules}
\end{center}

\begin{center}
{\large S. S. Afonin\footnote{Email: \texttt{s.afonin@spbu.ru}.}}
\end{center}

\begin{center}
{\it Saint Petersburg State University, 7/9 Universitetskaya nab.,
St.Petersburg, 199034, Russia}
\end{center}

\begin{abstract}
The calculation of the mass of light scalar isosinglet meson
within the Shifman--Vainshtein--Zakharov (SVZ) sum rules is
revisited. We develop simple analytical methods for estimation of
hadron masses in the SVZ approach and try to reveal the origin of
their numerical values. The calculations of hadron parameters in
the SVZ sum rules are known to be heavily based on a choice of the
perturbative threshold. This choice requires some important {\it
ad hoc} information. We show analytically that the scalar mass
under consideration has a lower and upper bound which are
independent of this choice: $0.78\lesssim m_s\lesssim1.28$~GeV. 
\end{abstract}

\section{Introduction}

The enigmatic $\sigma$-meson, called also $f_0(500)$ in the modern
Particle Data~\cite{pdg}, is attracting a lot of interest for a
long time. This lightest scalar resonance emerges usually as an
indispensable ingredient in description of the nuclear forces and
chiral symmetry breaking in the strong interactions (see, e.g.,
the recent review~\cite{pelaez}). The studies of this meson have a
rich and dramatic history~\cite{pelaez}. Recently the great
efforts have led to a significant progress in reducing the
uncertainty in its mass and full width~\cite{pdg}. One observes an
increasing evidence in favor of non-ordinary nature of this broad
resonance which hardly can be accommodated within the usual
quark-antiquark picture of mesons~\cite{pelaez}.

In the modern literature, the $\sigma$-meson is mainly studied in
the framework of methods based on analyticity and
unitarity~\cite{pelaez}. The ensuing models does not have direct
relations with QCD, may be except the studies of large-$N_c$
behavior of meson masses. The relations with QCD of many older
studies based on effective field theory, bag models, {\it
etc.}~\cite{pelaez} are also unclear. Among the phenomenological
approaches, the method that perhaps is mostly related to QCD in
the spectroscopy of light mesons represents the method of
Shifman--Vainshtein--Zakharov (SVZ) sum rules~\cite{svz}, often
called also ITEP or spectral or just QCD sum rules. The philosophy
of this approach is based on the assumption that a quark-antiquark
pair (or a more complicated quark current interpolating a hadron)
being injected in the strong QCD vacuum does not perturb it
noticeably. This allows to parametrize the unknown
non-perturbative vacuum by some universal phenomenological
characteristics --- the vacuum condensates. Hadrons with different
quantum numbers have different masses (decay constants,
formfactors, {\it etc.}) because their currents react differently
to the vacuum medium. And, roughly speaking, the corresponding
coefficients can be calculated from QCD. Assuming further the
existence of resonance in some energy range, one is able to
calculate its characteristics via the dispersion relations and the
Operator Product Expansion (OPE). If one of these assumptions
fails, the SVZ method should not work. In practice, this method
turned out to be extremely successful in description of hadron
parameters~\cite{colangelo}.

The SVZ sum rules predict that the expected mass of the lightest
scalar meson composed of quark and antiquark lies near
1~GeV~\cite{rry}. In order to obtain a smaller mass, say in the
interval 500--700 MeV, one typically needs to consider a
four-quark current (see a corresponding review in
Ref.~\cite{pelaez}). This looks like a confirmation of
non-ordinary nature of the $\sigma$-meson as long as the
assumption of its tetraquark structure works well in various
approaches~\cite{pelaez}.

However, the calculations of hadron masses in the SVZ method
involve rather strong assumptions, first of all about a choice of
the perturbative threshold. In view of a prominent role that the
$\sigma$-meson plays both in nuclear and particle physics, it is
highly desirable to reduce any {\it ad hoc} assumptions in studies
of this remarkable resonance as much as possible. The main purpose
of the present work is to demonstrate explicitly that the mass of
the quark-antiquark scalar meson in the SVZ sum rules have a lower
bound which is independent of the perturbative threshold and lies
above the expected $\sigma$-meson mass. As a by-product, we
develop a method of simple analytical estimations of hadron masses
within the QCD sum rules and demonstrate it in some important
cases. In particular, we derive an upper bound on the mass in
question.

In numerous papers devoted to the SVZ phenomenology, the values of
hadron parameters are customary obtained numerically and
demonstrated graphically. The style of our analysis is rather
unusual --- we will mainly follow various analytical estimates.

The paper is organized as follows. We recall briefly the SVZ
method in Section~2. In Section~3, this method is demonstrated in
the scalar case. We also explain how the obtained numerical result
can be simply calculated analytically. The lower bound on the
scalar mass is derived in Section~4. In Section~5, we obtain the
upper bound on this mass. The origin of numerical values of meson
masses in the SVZ sum rules is discussed in Section~6. We conclude
in Section~7. An application of some of our ideas to the
axial-vector case is demonstrated in the Appendix.

\section{SVZ method in the scalar sector}

We recall briefly the derivation of SVZ sum rules in the scalar
sector. Consider the two-point correlation function
\begin{equation}
\label{1}
\Pi_s(p^2)=i\int d^4x\, e^{ipx}\left\langle0|\text{T}\left\{j_s(x),j_s(0)\right\} \right|0\rangle.
\end{equation}
Here the scalar current is $j_s=\bar{q}q$, where the symbol $q$
stays for the $u$ or $d$ quark.
The object~\eqref{1} contains much dynamical information. In
particular, the large-distance asymptotics of~\eqref{1} in the
Euclidean space is $\sim e^{-m_s|x|}$, where $m_s$ is the mass of
the ground state, the scalar one in our case. This property lies
in the base of the lattice calculations of hadron masses directly
from QCD. The central problem in the classical SVZ sum rules
consists in the extraction of hadron masses from~\eqref{1} with
the help of some (semi)analytical methods and several
phenomenological inputs. The whole approach is based on the two
main ideas: The use of the Operator Product Expansion
for~\eqref{1} and a representation of the correlator~\eqref{1} via
a suitable dispersion relation.

In the Euclidean domain ($p^2=-Q^2$), the OPE for~\eqref{1}
reads~\cite{rry}
\begin{multline}
\label{2}
\Pi_s(Q^2)=\frac{3}{16\pi^2}\left(1+\frac{11}{3}\frac{\alpha_s}{\pi}\right)Q^2\log{\frac{Q^2}{\mu^2}}+
\frac{\alpha_s}{16\pi}\frac{\langle G^2\rangle}{Q^2}
+\frac32\frac{m_q\langle\bar{q}q\rangle}{Q^2}\\
-\frac{88}{27}\pi\alpha_s\frac{\langle\bar{q}q\rangle^2}{Q^4}+\mathcal{O}\left(\frac{1}{Q^6}\right),
\end{multline}
where $\langle G^2\rangle$ and $\langle\bar{q}q\rangle$ denote the
gluon and quark vacuum condensate, respectively. In the practical
calculations of masses for the light non-strange mesons, one
neglects: (i) The running of $\alpha_s$ and of the factor in front
of $\alpha_s$ in the unit operator; (ii) The $\mathcal{O}(m_q^2)$
and $\mathcal{O}(1/Q^6)$ contributions; (iii) A small anomalous
dimension of $\alpha_s\langle\bar{q}q\rangle^2$. The vacuum
saturation hypothesis~\cite{svz} is exploited for the
dimension-six operator in~\eqref{2} (i.e. the factorization
$\langle\bar{q}\Gamma q\bar{q}\Gamma
q\rangle\sim\left[(\text{Tr}\Gamma)^2-
(\text{Tr}\Gamma^2)\right]\langle\bar{q}q\rangle^2$ which can be
justified in the large-$N_c$ limit of QCD) and the value of
$\alpha_s\langle\bar{q}q\rangle^2$ absorbs small contribution from
other dimension-six operators ($m_q\langle\bar{q}Gq\rangle$ and
$\langle G^3\rangle$).

On the other hand, the correlation function~\eqref{1} satisfies
the twice-subtracted dispersion relation,
\begin{equation}
\label{3}
\frac{d^2\Pi_s(p^2)}{d(p^2)^2}=\frac{2}{\pi}\int_{4m_q^2}^{\infty} ds
\frac{\text{Im}\Pi_s(s)}{(s-p^2)^3}.
\end{equation}
In the SVZ sum rules, one usually assumes the so-called "one
resonance plus continuum" ansatz for the spectral function,
\begin{equation}
\label{4}
\text{Im}\Pi_s=f_s^2s\delta(s-m_s^2)+\frac{3}{16\pi^2}\left(1+\frac{11}{3}\frac{\alpha_s}{\pi}\right)s\Theta(s-s_0).
\end{equation}
Here the scalar "decay constant" is defined by the matrix element
of scalar current between the vacuum and a scalar state $f_0$,
\begin{equation}
\label{5}
\langle0|j_s|f_0\rangle=f_sm_s.
\end{equation}
The ansatz~\eqref{4} does not take into account the decay width,
i.e. the resonance is considered as infinitely narrow.
The higher resonances and various thresholds in the spectral
function are absorbed into the perturbative continuum.

In order to suppress the higher dimensional condensates in the OPE
and simultaneously enhance the relative contribution of the ground
state into the correlator, the SVZ method makes use of the Borel
transform,
\begin{equation}
\label{6}
L_M\Pi(Q^2)=\lim_{\substack{Q^2,n\rightarrow\infty\\Q^2/n=M^2}}\frac{1}{(n-1)!}(Q^2)^n\left(\frac{-d}{dQ^2}\right)^n\Pi(Q^2).
\end{equation}
The transform~\eqref{6} suppresses operators of dimension $2k$ by
a factor of $1/(k-1)!$ and provides an exponential suppression of
the kind of $e^{-m_n^2/M^2}$ for heavier resonances ("radial
excitations" with masses $m_n$) in the spectral
function~\eqref{4}. Applying~\eqref{6} to (twice derivative
of)~\eqref{2} and~\eqref{3} with ansatz~\eqref{4} one arrives at
the expression $f_s^2e^{-m_s^2/M^2}=...$. Another relation can be
obtained by considering the derivative
$\frac{d}{d(1/M^2)}(-f_s^2e^{-m_s^2/M^2})=f_s^2m_s^2e^{-m_s^2/M^2}=...$
(the ensuing relation represents the sum rule for derivative of
$Q^2\Pi_s(Q^2)$). Dividing the second relation by the first one,
we get finally~\cite{rry}
\begin{equation}
\label{7}
m_s^2=M^2\frac{2h_0\left[1-\left(1+\frac{s_0}{M^2}+\frac{s_0^2}{2M^4}\right)e^{-s_0/M^2}\right]+\frac{h_3}{M^6}}
{h_0\left[1-\left(1+\frac{s_0}{M^2}\right)e^{-s_0/M^2}\right]+\frac{h_2}{M^4}-\frac{h_3}{M^6}},
\end{equation}
where
\begin{eqnarray}
\label{8}
h_0&=&1+\frac{11}{3}\frac{\alpha_s}{\pi},\\
\label{9}
h_2&=&\frac{\pi^2}{3}\left(\frac{\alpha_s}{\pi}\langle G^2\rangle+24m_q\langle\bar{q}q\rangle\right),\\
\label{10}
h_3&=&\frac{1408}{81}\pi^3\alpha_s\langle\bar{q}q\rangle^2.
\end{eqnarray}
The numerical values taken in Ref.~\cite{rry} are: $\alpha_s=0.6$,
$h_2=0.04$~GeV$^4$, $h_3=0.08$~GeV$^6$.

\section{Calculation of the scalar mass}

The extraction of scalar mass from the expression~\eqref{7} is a
matter of art as it requires some additional information. First,
the result is sensitive to the choice of the continuum threshold
$s_0$. So {\it apriori} we need to guess the mass region where our
resonance should lie and we should have certain ideas on the mass
of the next scalar meson (the first "radial excitation") and on
positions of various thresholds in the spectral function. Second,
the prediction should be stable against variations of the borel
parameter $M$ in the so-called "Borel window" $a<M^2<b$ in which
we hope to find our resonance. In the region $M^2<a$, the power
corrections in OPE~\eqref{2} become important and we cannot
consider the first few terms only. In the region $M^2>b$, the
heavier resonances and various thresholds contribute to the
spectral function~\eqref{4}. After some discussions, the choice
made in Ref.~\cite{rry} was (in~GeV$^2$) $s_0\simeq1.5$ and
$0.8<M^2<1.4$ that resulted in the prediction
\begin{equation}
\label{11}
m_s=1.00\pm0.03\,\text{GeV}.
\end{equation}
The given estimate is usually regarded as a standard prediction
for the scalar mass in the SVZ sum rules.

We wish to make a comment on this result which will be important
in what follows. In the classical case of $\rho$-meson~\cite{svz},
the mass $m^2_\rho$ as a function of $M^2$ has a rather deep
minimum. The Borel window lies around this minimum near $M^2\simeq
m^2_\rho$ and the dependence of $m^2_\rho$ on $s_0$ is not strong.
The scalar case is different: The minimum of function $m_s^2(M^2)$
in~\eqref{7} is very shallow and hardly visible in the practical
calculations. In this case, the dependence on $s_0$ becomes much
stronger. Our key observation is that within the accuracy
displayed in~\eqref{11}, the value of $m_s^2$ predicted in the
Borel window practically coincides with its asymptotic value at
$M^2\rightarrow\infty$,
\begin{equation}
\label{12}
m^2_{\text{asymp}}=\frac{\frac13h_0s_0^3+h_3}{\frac12h_0s_0^2+h_2}.
\end{equation}
The asymptotics~\eqref{12} is independent of $\mathcal{O}(1/M^8)$
condensate contributions omitted in~\eqref{7} but deviations
of~\eqref{12} from $m_s^2$ calculated in the Borel window do
depend on these contributions. If we accept the accuracy 0.03~GeV
as in~\eqref{11}, we will have
$m_{\text{asymp}}-m_s\lesssim0.03$~GeV when
$s_0\lesssim1.7$~GeV$^2$. For $s_0\simeq1.5$~GeV$^2$ used
in~\eqref{11}, the deviation is
$m_{\text{asymp}}-m_s\simeq0.01$~GeV.

If $s_0$ in~\eqref{12} is large enough the condensate
contributions $h_2$ and $h_3$ are not substantial. In this
"physical" case, we obtain from~\eqref{12} a simple expression for
the scalar mass,
\begin{equation}
\label{13}
m^2_s\simeq\frac23s_0.
\end{equation}
Taking the input $s_0\simeq1.5$~GeV$^2$ used for the numerical
result~\eqref{11}, this result follows immediately. The
relation~\eqref{13} also explains why (as was observed in
Ref.~\cite{rry} numerically) $m_s$ is very insensitive to a change
in $\alpha_s$: The dependence on $\alpha_s$ appears only in the
next-to-leading level in the $\frac{1}{s_0}$ expansion,
$m^2_s\simeq\frac23s_0-\frac{4h_2}{3h_0s_0}$.

To justify further the usefulness of the limit
$M^2\rightarrow\infty$, in Appendix, we demonstrate how it works
in the axial-vector sector.

\section{Lower bound on the scalar mass}

The scalar mass $m_s$ depends rather strongly on the choice of
perturbative cutoff parameter $s_0$, however, there is a lower
bound on the value of $m_s$. Since in the region
$s_0\lesssim1.5$~GeV$^2$ the asymptotic formula~\eqref{12} yields
a result practically indistinguishable from the exact numerical
value within the accuracy of the SVZ sum rules, we first consider
this region.

This bound can be obtained from calculation of minimum of
$m_s^2(s_0)$ from~\eqref{12}. We have
\begin{equation}
\label{17}
\frac{dm_s^2}{ds_0}=\frac{h_0s_0}{\left(\frac12h_0s_0^2+h_2\right)^2}
\left[s_0\left(\frac12h_0s_0^2+h_2\right)-\left(\frac13h_0s_0^3+h_3\right)\right]=0.
\end{equation}
The equation~\eqref{17} is quartic in $s_0$. It has 4 solutions
and $s_0=0$ is one of them. Thus, there exists at least another
one real solution that we will denote $\bar{s}_0$. In fact, at
positive $h_2$ and $h_3$, the remaining two solutions are always
complex since the imaginary part of roots is $\pm
i\frac{\sqrt{3}}{2}\left(H+\frac{2h_2}{h_0H}\right)$, where $H$ is
given by~\eqref{23}. It is easy to check that
\begin{equation}
\label{18}
\left.\frac{d^2m_s^2}{ds_0^2}\right|_{s_0=0}=-\frac{h_0h_3}{h_2^2}<0.
\end{equation}
Hence, $s_0=0$ delivers the local maximum to~\eqref{12}. Thus,
$s_0=\bar{s}_0$ corresponds to the (global at $s_0>0$) minimum
of~\eqref{12}. To calculate $m_s^2(\bar{s}_0)$ we do not need the
exact expression for $\bar{s}_0$. Equating to zero the square
brackets in~\eqref{17} we get
\begin{equation}
\label{19}
\bar{s}_0=\frac{\frac13h_0\bar{s}_0^3+h_3}{\frac12h_0\bar{s}_0^2+h_2}.
\end{equation}
Comparing~\eqref{19} with~\eqref{12} we conclude immediately that
\begin{equation}
\label{20}
\left.m_s^2\right|_{\text{min}}=\bar{s}_0,
\end{equation}
where, from~\eqref{19}, $\bar{s}_0$ is the real solution of the cubic equation
\begin{equation}
\label{21}
\frac16h_0s_0^3+h_2s_0-h_3=0.
\end{equation}
This solution can be written explicitly,
\begin{equation}
\label{22}
\bar{s}_0=H-\frac{2h_2}{h_0H},
\end{equation}
\begin{equation}
\label{23}
H=\left(\frac{\sqrt{9h_3^2+8\frac{h_2^3}{h_0}}+3h_3}{h_0}\right)^{1/3}.
\end{equation}

For further numerical estimates, we explain in detail how the
input parameters determining $h_0$, $h_2$ and $h_3$
in~\eqref{8}--\eqref{10} are fixed. The GOR relation for the pion
mass, $m_\pi^2f_\pi^2=-(m_u+m_d)\langle\bar{q}q\rangle$ can be
directly derived in the SVZ sum rules~\cite{svz}. This allows to
fix the renorminvariant dim-4 condensate
$m_q\langle\bar{q}q\rangle=-\frac12m_\pi^2f_\pi^2$, where
$m_\pi=140$~MeV and $f_\pi=92.4$~MeV~\cite{pdg}. To fix
$\alpha_s\langle\bar{q}q\rangle^2$ we will exploit the approximate
renorminvariance of this dim-6 condensate. Since
$\left.\alpha_s\right|_{\mu=2\text{GeV}}\simeq0.3$ and
$\left.(m_u+m_d)\right|_{\mu=2\text{GeV}}\simeq7$~MeV~\cite{pdg},
we have
$\alpha_s\langle\bar{q}q\rangle^2\simeq0.3\frac{m_\pi^4f_\pi^4}{0.007^2}$~GeV$^6$.
We will use $\alpha_s$ at the scale 1~GeV where the scalar meson
is expected. As
$\left.m_q\right|_{\mu=1\text{GeV}}\simeq1.35\left.m_q\right|_{\mu=2\text{GeV}}$~\cite{pdg},
we get
$\left.\langle\bar{q}q\rangle\right|_{\mu=1\text{GeV}}\simeq1.35^{-1}
\left.\langle\bar{q}q\rangle\right|_{\mu=2\text{GeV}}$ and the
approximate renorminvariance of $\alpha_s\langle\bar{q}q\rangle^2$
leads to
$\left.\alpha_s\right|_{\mu=1\text{GeV}}\simeq1.35^2\left.\alpha_s\right|_{\mu=2\text{GeV}}\simeq0.55$.
The mostly used phenomenological value of the gluon condensate is
$\frac{\alpha_s}{\pi}\langle G^2\rangle\simeq
(0.36~\text{GeV})^4$.

With the input values above, the relations~\eqref{8}--\eqref{10}
fix $h_0\simeq1.64$, $h_2\simeq0.049~\text{GeV}^4$,
$h_3\simeq0.092~\text{GeV}^6$. Using these inputs, the
relations~\eqref{20}--\eqref{23} yield the following estimate:
$\left.m_s\right|_{\text{min}}\simeq0.78$~GeV. This lower bound
practically coincides with the $\omega$-meson mass~\cite{pdg}.
Taking $\alpha_s$ at the scale of the $\omega$-meson mass,
$\alpha_s\simeq0.7$ (as it was originally used in
Ref.~\cite{svz}), one gets a bit corrected estimate:
$\left.m_s\right|_{\text{min}}\simeq0.77$~GeV. The uncertainty in
determination of the gluon condensate,
$\frac{\alpha_s}{\pi}\langle G^2\rangle\simeq
(0.36\pm0.02~\text{GeV})^4$, leads to the uncertainty
$\left.m_s\right|_{\text{min}}\simeq0.77\pm0.01$~GeV.

A further correction may come from  a more accurate treatment of
the large-$N_c$ limit in QCD~\cite{hoof,wit}. Since this limit was
exploited to justify both the narrow-width approximation in the
spectral function~\eqref{4} and the vacuum saturation hypothesis
for the dim-6 condensate in the OPE~\eqref{2}, it looks more
consistent to use the large-$N_c$ limit also for the factor in
front of the dim-6 operator in the OPE~\eqref{2}:
$\frac{88}{27}=\frac{11}{9}\left(N_c-\frac{1}{N_c}\right)$. This
point is usually ignored in the SVZ phenomenology. Neglecting the
$1/N_c$ term, i.e. replacing
$\frac{88}{27}\rightarrow\frac{11}{3}$, slightly shifts up our
estimate:
$\left.m_s\right|_{\text{min}}\simeq0.77\rightarrow0.79$~GeV.

In summary, we would estimate the lower bound of the scalar mass
as
\begin{equation}
\label{24}
\left.m_s\right|_{\text{min}}=0.78\pm0.02\,\text{GeV}.
\end{equation}
It is seen that at physical values of condensates the lower bound
for the mass of scalar meson lies around the mass of the
$\omega$-meson. If we imposed by hands a typical $\sigma$-meson
mass, say $m_\sigma\simeq0.45$~GeV~\cite{pdg,pelaez}, as the lower
bound at physical $\langle\bar{q}q\rangle$, we would obtain that
this is possible at a huge value of the gluon condensate,
$\frac{\alpha_s}{\pi}\langle G^2\rangle\simeq
(0.61~\text{GeV})^4$. Alternatively, we could fix the physical
value for $\frac{\alpha_s}{\pi}\langle G^2\rangle$ and obtain then
a rather small quark condensate
$\left.\langle\bar{q}q\rangle\right|_{\mu=1\text{GeV}}\simeq-(0.19~\text{GeV})^3$
(its physical value lies around $-(0.25~\text{MeV})^3$ at
$\mu=1$~GeV).

Consider now the region $s_0\gtrsim1.5$~GeV$^2$. Here the
deviation of calculated mass from the asymptotics~\eqref{12}
becomes visible. However, the predicted value of the scalar mass
exceeds the lower bound~\eqref{24} significantly,
$m_s\gtrsim1$~GeV, and grows with $s_0$. Thus, this region does
not change our estimate~\eqref{24}.

\section{Upper bound on the scalar mass}

If the perturbative threshold $s_0$ is high enough, say
$s_0\gtrsim2$~GeV$^2$, the minimum of the function $m_s^2(M^2)$
in~\eqref{7} becomes prominent. The Borel window is situated
around this minimum. When $s_0$ grows, the competition between the
perturbative continuum and power corrections narrows the Borel
window. In the limit $s_0\rightarrow\infty$, the Borel window
shrinks to one "fixed point" $M_0$ which represents the minimum of
$m_s^2(M^2)$. The scalar mass, as a quantity calculated in the
Borel window, grows with $s_0$ and reaches its maximal value in
this point, $\left.m_s\right|_{\text{max}}=m_s(M_0^2)$.

Regarding the limit $s_0\rightarrow\infty$, one may wonder why we
may neglect the contribution of thresholds and higher resonances
to the spectral function~\eqref{4}? First of all, the
ansatz~\eqref{4} is rough as the decay width is neglected. The
narrow-width approximation is justified in the large-$N_c$ limit
of QCD~\cite{hoof,wit}. All thresholds related with particle decays
are suppressed in this limit and should be then also neglected. If
an analysis is based on the large-$N_c$ arguments, one cannot
pretend to the accuracy better than, say, 10\%. In practice, the
contribution of heavier resonances (the radial excitations) to the
spectral function is rather small and comparable with 10\%. Thus a
possible error from omitting the radial states lies within the
accuracy of the method itself.

With this caveat in mind, consider the limit
$s_0\rightarrow\infty$ in~\eqref{7},
\begin{equation}
\label{25}
m^2_{s_0\rightarrow\infty}=M^2\frac{2h_0+\frac{h_3}{M^6}}{h_0+\frac{h_2}{M^4}-\frac{h_3}{M^6}}.
\end{equation}
The minimum of expression~\eqref{25} follows from the condition
$\frac{dm^2_{s_0\rightarrow\infty}}{dM^2}=0$ that leads to the
polynomial equation
\begin{equation}
\label{26}
2h_0^2(M^2)^6+6h_0h_2(M^2)^4-10h_0h_3(M^2)^3-h_3^2=0.
\end{equation}
With our numerical inputs above, the positive real solution of~\eqref{26} is
\begin{equation}
\label{27}
M_0\simeq0.78\,\text{GeV}.
\end{equation}
Substituting this value to~\eqref{25} we conclude that the
calculation of the scalar mass in the Borel window cannot exceed
the upper limit
\begin{equation}
\label{28}
\left.m_s\right|_{\text{max}}\simeq1.28\,\text{GeV}.
\end{equation}
The upper bound~\eqref{28} practically coincides with the mass of
$f_1(1285)$-meson~\cite{pdg} which, like a usual scalar meson, is
also a $P$-wave quark-antiquark state.

\section{Discussions}

It is interesting to observe a numerical coincidence of the Borel
"fixed point" $M_0$ in~\eqref{27} and the lower bound for the
scalar mass~\eqref{24}. This coincidence is not completely
accidental. The value of $h_3^2$ in the Eq.~\eqref{26} is
numerically very small, omitting this term does not affect the
result~\eqref{27} within our accuracy. The Eq.~\eqref{26} is then
cubic in $M^2$ and can be solved analytically. The most important
point here is that the relative contribution of the term with
$h_2$ into the real solution of both Eq.~\eqref{26} and
Eq.~\eqref{21} is relatively small (numerically by almost two
orders of magnitude) in comparison with the term containing $h_3$.
This allows to write immediately the approximate real solution of
Eq.~\eqref{21}, $\sqrt{\bar{s}_0}\simeq(6h_3/h_0)^{1/6}$, and of
Eq.~\eqref{26}, $M_0\simeq(5h_3/h_0)^{1/6}$. the relative
difference is $(6/5)^{1/6}\simeq1.03$ and explains qualitatively
the numerical coincidence of $M_0$ with
$\left.m_s\right|_{\text{min}}$. In addition, the given
observation demonstrates that both $\left.m_s\right|_{\text{min}}$
and (via $M_0$) $\left.m_s\right|_{\text{max}}$ are determined
mainly by the value of the condensate of the highest dimension
kept in the OPE.

Physically $M_0$ can be interpreted as the parameter determining
the left border of the Borel window. When the perturbative
threshold is moved from infinity to some finite physical value,
say around $s_0\simeq1.5$~GeV$^2$, the left border slightly shifts
right (enlarges) from $M_0$. As a consequence, the value of mass
calculated in the Borel window slightly moves down from the
maximal bound, in the scalar case --- the expression~\eqref{25}.
Here the term "slightly" means "within 10--15\%", i.e. within the
accuracy of the SVZ method. The seemingly strong dependence of
mass on $s_0$, expressed by the relation~\eqref{13}, appears only
in a very limited interval of $s_0$. In reality, however, the
extracted mass is mainly determined by the value of $M_0$. Indeed,
this value has an intermediate position between the physical
cutoff $\sqrt{s_0}$ and (a proper power of) condensate
contributions. This allows to write a simple estimate for the
scalar mass just neglecting the exponential and power terms
in~\eqref{7} (they are suppressed simultaneously),
\begin{equation}
\label{29}
m_s^2\simeq 2M_0^2.
\end{equation}
This estimate yields $m_s\simeq1.1$~GeV that agrees within 10\%
with~\eqref{11}.

Consider the classical case of $\rho$-meson. The mass relation
is~\cite{svz,rry}
\begin{equation}
\label{30}
m_\rho^2=M^2\frac{h_0\left[1-\left(1+\frac{s_0}{M^2}\right)e^{-s_0/M^2}\right]-\frac{h_2}{M^4}-\frac{h_3}{M^6}}
{h_0\left[1-e^{-s_0/M^2}\right]+\frac{h_2}{M^4}+\frac{h_3}{2M^6}},
\end{equation}
where $h_0=1+\frac{\alpha_s}{\pi}$ with $\alpha_s=0.7$,
$h_2=0.046~\text{GeV}^4$, $h_3\simeq-0.064~\text{GeV}^6$.
Following the same consideration, we would find
$M_0\simeq0.72$~GeV and
\begin{equation}
\label{31}
m_\rho\simeq M_0.
\end{equation}
The given estimate justifies the assumption $m_\rho\simeq M$ used
in the original paper~\cite{svz}, where $M$ was an "optimal"
region in the Borel window.

In the classical SVZ sum rules, one usually estimates the
contribution of condensates to the masses on the level of
5--10\%~\cite{svz}. We can trace qualitatively how this estimate
originates within our approach: Roughly speaking, the estimate
stems from neglecting the power corrections in~\eqref{31}
or~\eqref{29}. However, we can directly see that this estimate is
not well justified because the value of $M_0$ is completely
determined by the condensates, as was shown above. To put it
differently, the scale of a meson mass extracted from the SVZ sum
rules is mainly dictated not by the perturbative threshold $s_0$
(as it might seem) but by the scale of the Borel parameter $M$ in
the Borel window. And this latter is determined (via $M_0$) by the
condensates!

Recalling that the numerical value of $M_0$ comes mainly from the
condensate of the highest dimension kept in the OPE, it looks like
a miracle that the SVZ sum rules work well neglecting all higher
power corrections. A partial reason might be the fact that the
values of condensates in the SVZ method are normalized to the
phenomenology and an account for higher power terms could lead to
a double counting of non-perturbative effects. But the main reason
of phenomenological success of the SVZ method seems to consist in
a fortunate parametrization of important universal properties of
the non-perturbative QCD vacuum.

It should be emphasized that we have used the simplest SVZ sum rules 
framework, i.e. we employed in the scalar sector the same set of 
assumptions as in the vector one. Our wish was to escape any
additional model assumptions and related complications. In reality,
the scalar sector is known to be much more complicated than the 
vector one. A more advanced study of the correlator of the scalar
currents should include, e.g. higher order perturbative contributions,
instanton contributions, finite width effects, and mixing with glueballs.
An example of such an analysis for the non-strange quark-antiquark
scalar sector is given in Refs.~\cite{steele1,steele2}.

\section{Conclusions}

Within the framework of SVZ sum rules, we have shown that the mass
of scalar resonance interpolated by the quark-antiquark current
has a lower and upper bound, $0.78\lesssim m_s\lesssim1.28$~GeV.
These bounds do not depend on the  choice of the perturbative
threshold. The values of bounds are determined by the dim-4 and
dim-6 condensates in the OPE, with the main contribution stemming
from the dim-6 one.

Our analysis confirms a widespread idea that the $f_0(500)$-meson
represents an exotic state. The scalar isoscalar state described
by the SVZ method can physically correspond to $f_0(980)$ or
$f_0(1370)$. The latter possibility is much less likely as the
mass uncertainty of $f_0(1370)$, 1200--1500~MeV~\cite{pdg}, has a
relatively small overlap with the upper bound.

Our estimations can be extended to the scalar strange current
$j=\bar{s}s$. The bounds are expected to shift up by several
hundreds MeV. The (poorly known) higher dimensional condensates
should affect our analysis. The account for their effects is an
open problem.

\section*{Acknowledgments}

The work was supported by the Saint Petersburg State University
research grant 11.38.189.2014 and by the RFBR grant 16-02-00348-a.

\section*{Appendix}

The limit $M^2\rightarrow\infty$ is useful in the cases where a
hadron mass as a function of the Borel parameter $M$ does not have
a prominent extremum. In the given Appendix, we show how the
axial-vector sector is simplified in this limit.

Since the axial-vector current
$j_{\mu}^a=\bar{q}\gamma_\mu\gamma_5q$ is not conserved, the
two-point correlator~\eqref{1} of these currents has (in contrast
to the vector case) two independent contributions,
\begin{equation}
\label{A1}
\Pi_{\mu\nu}^a(p^2)=-\Pi_1(p^2)g_{\mu\nu}+\Pi_2(p^2)p_\mu p_\nu.
\end{equation}
The sum rules for $\Pi_1$ and $\Pi_2$ are different with different
final expressions for the axial-vector mass. The case of $\Pi_1$
is similar to the scalar one and the resulting mass formula is in
one-to-one correspondence with~\eqref{7}~\cite{rry},
\begin{equation}
\label{A2}
m_a^2=M^2\frac{2h_0\left[1-\left(1+\frac{s_0}{M^2}+\frac{s_0^2}{2M^4}\right)e^{-s_0/M^2}\right]+\frac{h_3}{M^6}}
{h_0\left[1-\left(1+\frac{s_0}{M^2}\right)e^{-s_0/M^2}\right]-\frac{h_2}{M^4}-\frac{h_3}{M^6}}.
\end{equation}
We will use the numerical values of parameters from
Ref.~\cite{rry}: $h_0=1+\frac{\alpha_s}{\pi}$ with $\alpha_s=0.6$,
$h_2=0.046~\text{GeV}^4$, $h_3\simeq0.10~\text{GeV}^6$. The
longitudinal part $\Pi_2$ contains a contribution from the pion
pole which enters the expression for the mass via the power term
with the factor
$h_1=8\pi^2f_\pi^2\simeq0.67~\text{GeV}^2$~\cite{svz,rry},
\begin{equation}
\label{A3}
m_a^2=M^2\frac{2h_0\left[1-\left(1+\frac{s_0}{M^2}\right)e^{-s_0/M^2}\right]-\frac{h_2}{M^4}-\frac{h_3}{M^6}}
{h_0\left[1-e^{-s_0/M^2}\right]-\frac{h_1}{M^2}+\frac{h_2}{M^4}+\frac{h_3}{2M^6}}.
\end{equation}
The numerical calculations show that the expressions~\eqref{A2}
and~\eqref{A3} result in the same mass for $s_0\simeq1.75$~GeV$^2$
which gives the mass
\begin{equation}
\label{A3b}
m_a=1.15\pm0.04\,\text{GeV}.
\end{equation}

As in the relation~\eqref{29}, we can write a simple estimate
$m_a^2\simeq 2M_0^2$. Here the value of $M_0$ is determined by the
Eq.~\eqref{26} with the opposite sign for the term containing
$h_2$. The solution is $M_0\simeq0.9$~GeV, where the numerical
difference with~\eqref{27} comes mainly from a different $h_0$. We
have thus the estimate $m_a\simeq1.27$~GeV that agrees within 10\%
with~\eqref{A3b}.

Consider now the limit $M^2\rightarrow\infty$ in the
expressions~\eqref{A2} and~\eqref{A3}. We get the lower asymptotics
\begin{equation}
\label{A4}
m_{\text{asymp}}^2=\frac{\frac13h_0s_0^3+h_3}{\frac12h_0s_0^2-h_2},
\end{equation}
for~\eqref{A2} and the upper asymptotics
\begin{equation}
\label{A5}
m_{\text{asymp}}^2=\frac{\frac12h_0s_0^2-h_2}{h_0s_0-h_1},
\end{equation}
for~\eqref{A3}. Equating~\eqref{A4} with~\eqref{A5}, one obtains a
quartic polynomial equation for $s_0$ which has analytical
solutions. The result is $s_0\simeq1.79$~GeV$^2$ and
$m_a\simeq1.13$~GeV. It is seen that the difference with the exact
numerical solution of Ref.~\cite{rry} is small.

As in the scalar case, we can try to neglect the contribution of
condensates in~\eqref{A4} and get a simple approximate
relation~\eqref{13} for the axial mass,
\begin{equation}
\label{A6}
m_a^2\simeq\frac23s_0.
\end{equation}
This would lead to the estimate $m_a\simeq1.09$~GeV differing from
the numerical solution by less than 10\%.

In the asymptotical expression~\eqref{A5}, the limit
$s_0\rightarrow\infty$ for the mass estimation is bad since the
contribution from $h_1$ is not relatively small. Let us
expand~\eqref{A5} in $\frac{1}{s_0}$ up to the next-to-leading term,
\begin{equation}
\label{A7}
m_a^2\simeq\frac12s_0+\frac{h_1}{2h_0}.
\end{equation}
Equating~\eqref{A6} with~\eqref{A7} we arrive at the estimate
\begin{equation}
\label{A8}
s_0\simeq\frac{3h_1}{h_0}=\frac{24\pi^2f_\pi^2}{1+\alpha_s/\pi}\simeq1.7~\text{GeV}^2,
\end{equation}
which is very close to the numerical value used in~\eqref{A3b}.
Substituting~\eqref{A8} in~\eqref{A6}, we can express the
axial-vector mass via $f_\pi$,
\begin{equation}
\label{A9}
m_a^2\simeq\frac{16\pi^2f_\pi^2}{1+\alpha_s/\pi}.
\end{equation}
As was advocated in the original Ref.~\cite{svz} on the SVZ sum
rules, saturation of spectral function
$\frac{1}{\pi}\text{Im}\Pi_2$ by only the pion pole below the
$\rho$-meson mass leads to a successful relation
$m_\rho^2\simeq\frac{8\pi^2f_\pi^2}{1+\alpha_s/\pi}$. Combining
this relation with~\eqref{A9} we re-derive the famous Weinberg
relation~\cite{wein}, $m_a^2\simeq2m_\rho^2$.

\end{document}